\documentclass[runningheads]{llncs}
\usepackage[T1]{fontenc}
\usepackage{graphicx}

\usepackage{color}
\usepackage{booktabs}
\usepackage{amsmath}
\usepackage{hyperref}
\usepackage{amssymb}

\begin{document}
\title{ApplE: A Modular Ontology of Applied Ethics and Event Context for Ethical Decision Modeling}
\titlerunning{ApplE: Ontology of Applied Ethics and Event Context}
%
\author{Aisha Aijaz\inst{1}\orcidID{0000-0002-8137-106X}, Raghava Mutharaju\inst{2}\orcidID{0000-0003-2421-3935}, Manohar Kumar\inst{3}\orcidID{0000-0003-2540-1356}}
\authorrunning{A. Aijaz et al.}
%
\institute{Department of Computer Science Engineering, IIIT Delhi, New Delhi, India
\email{aishaa@iiitd.ac.in} 
\and
Mehta Family School for Data Science and AI, IIT Palakkad, Kerala, India
\email{raghava@iitpkd.ac.in} 
\and
Department of Social Sciences and Humanities, IIIT Delhi, New Delhi, India
\email{manohar.kumar@iiitd.ac.in}}

\maketitle              
\begin{abstract}
Applied ethics applies ethical decision-making to domain-specific contexts using contextual information such as agents, actions, temporal and spatial settings, and theoretical constructs such as utility, virtues, rights, and duties. However, representing an ethical decision is challenging as it may be abstract, context-sensitive, and semantically heterogeneous. Nevertheless, important ethical and contextual factors can be formally modeled to support structured ethical reasoning. Knowledge representation and reasoning provide a mechanism to translate abstract ethical concepts into machine-interpretable conceptual structures in the context of an event. To achieve this, we propose ApplE, an Applied Ethics ontology that models ethical theory and event context within a unified and modular conceptual framework for ethical decision-making. The ontology was developed using a modified version of the Simplified Agile Methodology for Ontology Development (SAMOD), which facilitates iterative refinement of classes and relationships, as well as the participation of a domain expert. The modular development of ApplE combines Ethics Theory with Event Context to capture semantic relationships between ethical principles, agents, actions, consequences, intentions, and domains. Using ApplE, we modeled a use case from the medical domain to demonstrate the ontology’s representational expressivity and reasoning capabilities. In addition to ontological reasoning and consistency checks, ApplE is also evaluated using the three-fold testing process of SAMOD. ApplE follows the FAIR principles and is positioned to be used as a reusable semantic and conceptual modeling resource for ethical AI systems and ontology-driven applications. All resources, including the complete ApplE ontology, competency questions, SPARQL and DL queries, SWRL rules, documentation, and tutorials, are available at \url{https://github.com/kracr/applied-ethics-ontology/}.

\end{abstract}

\section{Introduction}
\label{Introduction}

Significant efforts in the past few decades have been made to integrate intellectual, emotional, and even physical capabilities into AI systems through powerful learning algorithms, natural language processing techniques, and advanced robotics \cite{zhong2025computational}. However, even state-of-the-art systems of today may not be considered inherently moral \cite{brozek2019can}. Considering how commonplace they are,  especially in critical settings, developing ethical machines becomes imperative. More so, given that they may interact with agents who may demand immoral action from them \cite{liao2020ethics}. An ethics taxonomy for AI would help them navigate ethical decision-making when confronted with difficult cases with real societal impact.

Categorizations have been presented \cite{moor2006nature} which generally consider moral machines as moral \textit{agents} or moral \textit{patients}. The former category places machines within the sphere of moral understanding and moral action, allowing them to affect other agents via \textit{agency}. The latter has to do with \textit{experience}, that of receiving an action and its consequences. This classification, although not concrete, is in accordance with moral psychology, specifically mind perception \cite{wegner2017mind}. The way we perceive our environment, agents involved, their intentions, predefined principles, and consequences of possible courses of action allow us to make ethical decisions. In other words, our actions are governed by our understanding of the ethical background and the event context.

We aim to develop a representation of applied ethics theory and event context that would be viable for autonomous ethical decision-making. By Moor's definition, an AI built upon this representation would be an \textit{explicit ethical agent} \cite{moor2006nature}, encoded with ethics to make ethically-informed decisions. Earlier implementations of ethics in machines by explicit discretization leverage concepts of deontology (driven by rights and duties), epistemology (requiring information or context), and action \cite{van2002deontic}. More recent embeddings of ethics in AI systems may involve consequentialism (by calculating utility) and virtue ethics (upholding ethical principles of virtue), as proposed by Greene et al. \cite{greene2016embedding}.

Determining an appropriate ethical course of action within a given context is challenging. It involves understanding factors sensitive to context, such as features, its constraints, and the circumstances that frame or limit the possibilities of action. It also involves many factors that are not explicit, such as the motivations of the agent. The complexity of arriving at an ethical decision is further complicated by the disagreement among different ethical theories, given a situation \cite{bergmann2014challenges}. For example, Immanuel Kant, a prolific philosopher in the area of ethics, would consider killing to be a punishable offense under a universal moral law, regardless of the context \cite{hill2013kantianism}, whereas classical utilitarians such as Jeremy Bentham and John Stuart Mill \cite{sep-utilitarianism-history} would allow it if doing so saves more lives than those being sacrificed. Similarly, virtue ethics considers honesty to be a trait of a virtuous, hence, ethical human being \cite{wilson2018honesty}. However, deontology pushes agents to follow the law and their duty as citizens to not spill state secrets \cite{mokrosinska2023necessary}. Thus, we notice pervasive disagreements among moral philosophers. 

Despite these challenges and acknowledgment of limitations, there have been successful implementations of ethics in AI systems, albeit in narrower scopes \cite{awad2018moral,anderson2018geneth,wu2018low}. After an extensive review of this research area, we have identified a lack of explicit implementation of applied ethics as a structured knowledge resource for AI systems. Consequently, we propose a formal ontology to capture a taxonomy of applied ethics. The aim of such a resource would be to take into account the various aspects of ethical decision-making in different domains such as business, healthcare, education, and the environment. This resource allows the development of a domain-agnostic, queryable knowledge graph, for use by an algorithmic notation of ethical judgement to arrive at an independent ethical decision \cite{aijaz2025moral}.

This paper makes the following contributions:
\begin{enumerate}
    \item A formal ontology for Applied Ethics and Event Context (ApplE).
    \item A real-world motivating scenario from bioethics modeled using ApplE, in accordance to the Spiral-SAMOD methodology.
    \item SWRL rules to indicate the moral nature of an action based on available real-world semantics.
\end{enumerate}

In Section \ref{Related work}, we discuss the gaps in the current literature and how our contributions may address them. The modified version of the SAMOD methodology \cite{peroni2016samod}, and the ontology development process are described in Section \ref{Methodology}. A description of ApplE's major modules has been outlined in Section \ref{OntologyDescription}. Technical details about ApplE as a resource are provided in Section \ref{Resource Specifications}. We evaluate the ontology using standard techniques and discuss a real-world motivating scenario in Section \ref{Evaluation}.

\section{Related Work}
\label{Related work}

Most applications of modeling ethics in computer systems refer to normative theories: utilitarianism, virtue ethics, and deontology. One or more of these theories are applied to specific use cases in computer applications, such as autonomous vehicles \cite{awad2018moral}, resource allocation \cite{10.1093/jlb/lsac012}, and racial discrimination \cite{briggs2020mitigating}.

Although non-symbolic techniques like learning algorithms \cite{wu2018low}, informal logic \cite{anderson2018geneth}, and empirical studies \cite{awad2018moral} have been used to model ethics in AI, they rely on data and frameworks rather than a structured understanding of ethical theory and its application. For example, the Moral Machine Project by MIT uses crowd-sourced data to model the Trolley Problem \cite{awad2018moral}. However, such data does not reliably reflect practical ethical reasoning. 

An explicit taxonomy of applied ethics is required to serve as an invariable source of ethics theory for computer systems to refer to and infer from when given certain context. This requirement may be fulfilled by a symbolic representation, or an ontology. 

\subsection{Existing Symbolic Representations of Related Concepts}

Lewis et al.'s ontology to standardize trustworthy AI by capturing international policies and regulations \cite{Lewis2021} provides some classes similar to what we have used for ApplE, however, the objective of the prior ontology is to provide structure to existing policy documents and does not cater to any explicit ethical decision-making. Similarly, the ontology presented by Vasquez et al. \cite{vasquez} judges emerging technologies based on their \textit{ethicality}, expectations, and social usability. Again, this does not aid the objective of developing a general taxonomy for applying ethics for dilemma resolution. DeBellis \cite{debellis2018universal} implements a Universal Moral Grammar (UMG) via ontology modeling, which operates on the premise of universal morality for all agents. This premise is more philosophical than practical, as it does not involve the contextual cues that are vital to applying ethics theory.

Guizzardi et al. \cite{guizzardi2023ontology} develop an ontology for ethicality requirements in engineering, similar to \cite{vasquez}, while also pertaining to intentions, values, and decisions. These are overlapping concepts that demonstrate the use of foundational ontological analysis to clarify semantics with regard to ethics. Similar concepts are also discussed by Sales et al. \cite{sales2018common}, who describe the Common Ontology of Value and Risk (COVER), describing explicit value and risk along with concepts such as agents, events, and context. Both these ontologies have some overlapping concepts with ApplE, however, they lack specific representation of theories in conjunction to context for an action. Our modular conceptualization aims to support and represent actual ethical decision-making across multiple domains.

The Deontic Cognitive Event Ontology \cite{vacura2020modeling}, models agents, actions, intentions, etc. within autonomous systems. It is grounded in deontic states and also supports context, but mainly focuses on the constraints governing these systems. This, too, is not a broad evaluation of the ethics of an action. ApplE is not confined to one normative or applied ethics theory or type of agent. 

\subsection{Research Gaps}
Overall, the landscape of related work reveals three main gaps in existing symbolic approaches to ethical reasoning. (a) Current ontologies are designed for specific applications instead of a general, reusable conceptualization of applied ethics across domains; (b) most approaches model ethical theory or contextual information, but do not generally integrate them for ethical deliberation; and (c) existing representations lack explanations and real-world semantic grounding to explain how the various classes such as theories, agents, and principles interact to produce an ethical assessment. ApplE addresses these gaps through modular development of Ethics Theory and Event Context modules to provide unified conceptualization and support for ethical deliberation across multiple domains.  

\section{Methodology}
\label{Methodology}

The ApplE ontology was developed primarily in accordance with the Simplified Agile Methodology for Ontology Development (SAMOD) \cite{peroni2016samod} and the Spiral Software Development Methodology \cite{boehm1988spiral}. This custom methodology supports quick, iterative cycles of development catered towards ontologies with fewer axioms. SAMOD describes the involvement of a domain expert (DE), in our case, an applied ethicist, who outlined the classes to be considered for the ontology in a taxonomy (see Section \ref{OntologyDescription}). The ontology engineers (OEs) implemented these entities. Although SAMOD recommends the involvement of the DE to be limited to the initial development phase, we deviated from this by involving the DE twice: once at the beginning to develop the initial modelet\footnote{A modelet is defined as a stand-alone model that describes a particular domain, which in our case would be applied ethics \cite{peroni2016samod}.} and then to model the motivating scenario described in Section \ref{Evaluation}. The requirement of continuous involvement of the domain expert is the main reason to not have used SAMOD directly. Additionally, the Spiral-SAMOD further simplified the flow of our methodology.  

\begin{figure*}[!ht]
\centering
\includegraphics[width=4in]{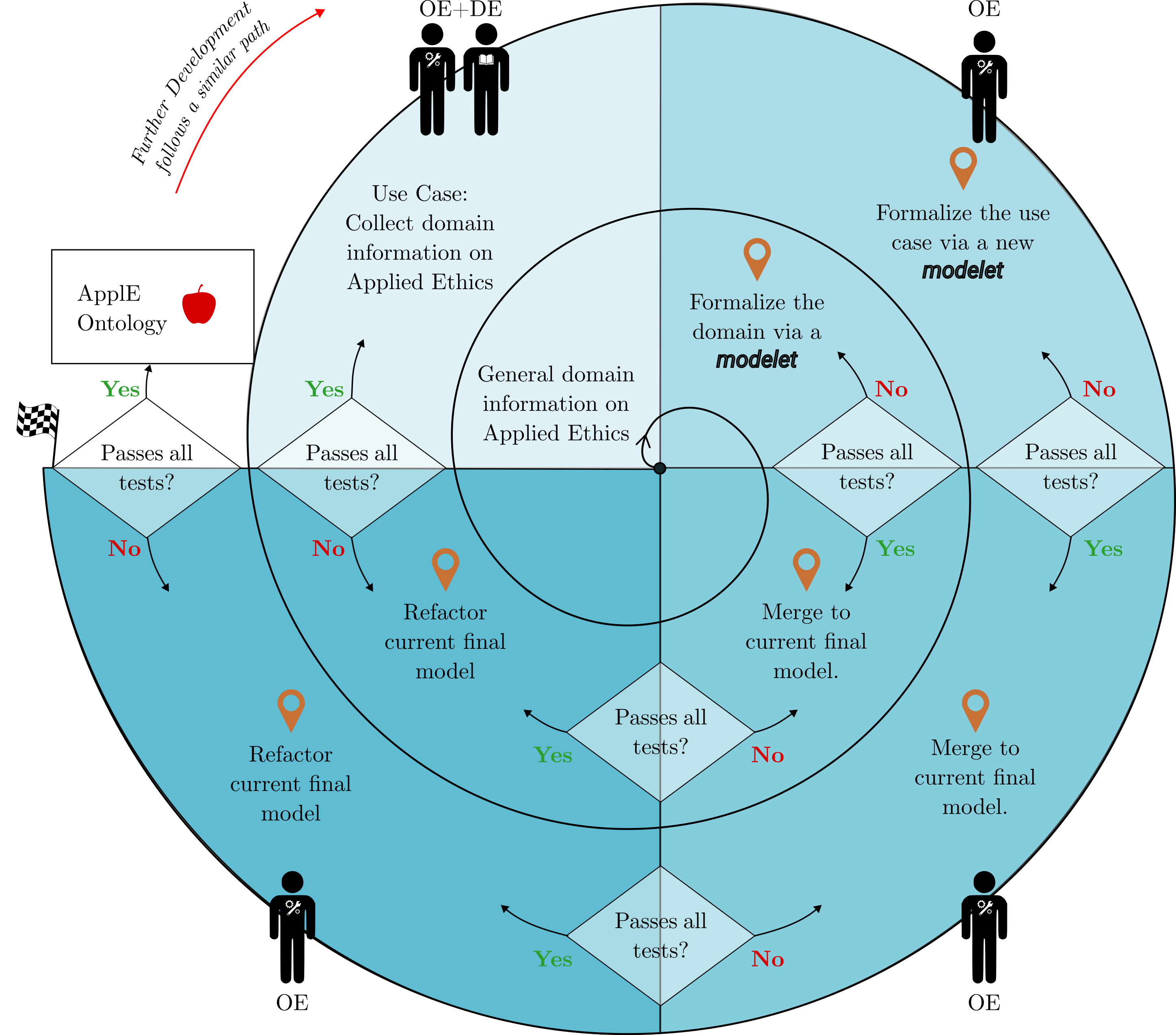}
\caption{Spiral-SAMOD: A representation of the hybrid methodology inspired by the Spiral methodology at a macro-level and the SAMOD methodology at a micro-level. Development starts at the center and ends as per requirement when one pass is complete. In our case, we underwent two passes of the methodology, as shown.} 
\label{methodology diagram}
\end{figure*}

Our methodology (Figure \ref{methodology diagram}) shows the four major quadrants of development:
\begin{enumerate}
    \item Gathering the domain information with the involvement of the DE. In the second pass, we developed the use cases.
    \item Formalizing the domain in a \textit{modelet}.
    \item Merge the modelet to the previous milestone.
    \item Refactor the new ontology to ensure semantic integrity.
\end{enumerate}

We selected the most relevant information, keeping ApplE small yet specific. We have also utilized multiple ontology design patterns (ODPs) and entities from established ontologies where possible. The ontology was built using middle-out expansion \cite{peroni2016samod}, and the naming scheme aims to be self-explanatory, thus aiding its value as a resource.

\begin{figure*}[!ht]
    \centering
    \includegraphics[width=\textwidth]{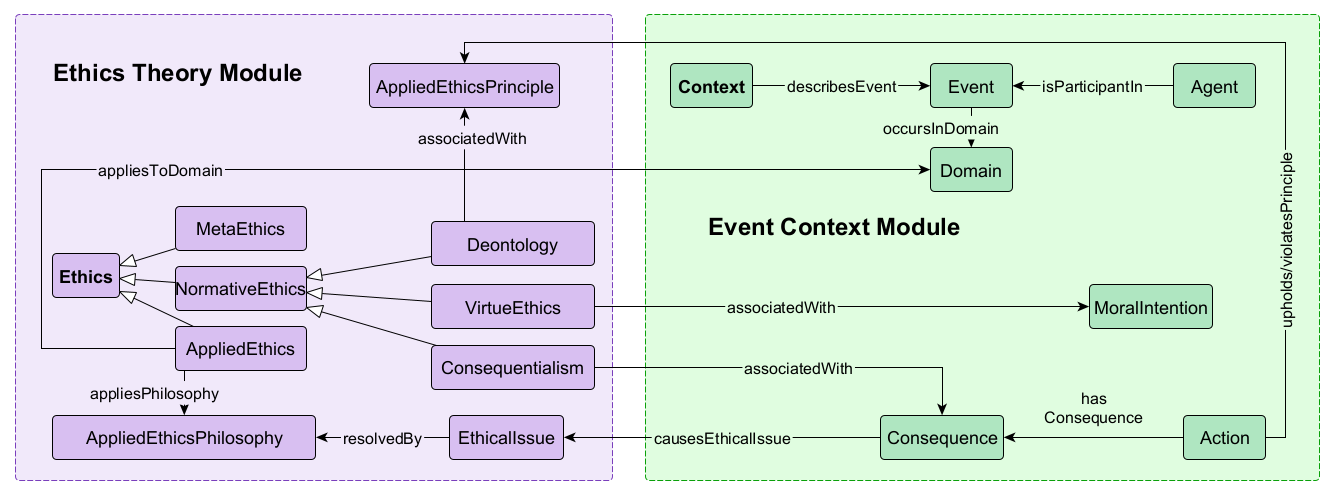}
    \caption{A high-level overview of the ApplE ontology with the two interacting modules: ethical theory and contextual parameters. These modules have been expanded in Section \ref{OntologyDescription}.}
    \label{upper-level ontology diagram}
\end{figure*}

Contextual (colored green in Figure \ref{upper-level ontology diagram}) and theoretical aspects (colored purple in Figure. \ref{upper-level ontology diagram}) are considered when making an ethical decision. The diagram shows the interacting concepts between the two categories, modeled separately in the ApplE ontology, to provide modular development. The relationships between these serve as bridge principles usually used in applied ethics to translate the theoretical normative philosophies into applicable concepts in real-world use cases \cite{nagel1990bridging}. This upper ontology is the first \textit{modelet} we created, which we expanded iteratively. 

The methodology also ensures \textit{modularity}, which allows for independent extension of subsets of the ontology based on required application without modifying ApplE's core structure. It can also be expanded with new ethical theories and domain knowledge. Our adherance to FAIR principles supports \textit{scalability} in parallel. ApplE is implemented in OWL2 DL, extended with Horn clause SWRL rules to enable \textit{expressive representation} of all pertinent components of an ethical deliberation. However, even with our focus on expressivity, we do not claim \textit{completeness} with regard to ethical philosophy. ApplE captures the core concepts required for applied ethical decision-making, with the possibility of independent modular expansion. 

ApplE explicitly accommodates for situations with conflicting moral conclusions and ethical principles through representation and interaction between multiple ethical criterion. Therefore, the indicative verdict on the morality of an action does not stem from a single parameter, it is dependent on a weighing of factuals and counterfactuals, tradeoffs, and available information.

The ontology was built in Protégé~\cite{DBLP:journals/aimatters/Musen15}. Various ontology design patterns and concepts were reused, and competency questions were developed for their evaluation \cite{aijaz2025appleappliedethicsontology}. After developing the motivating scenario and informal competency questions, the OEs and DEs put together a glossary of terms (Table \ref{got table}) that may be useful to the readers of this ontology. Finally, the three-fold testing was completed (see Section \ref{Evaluation}).

\section{Description of the Ontology}
\label{OntologyDescription}
After considering philosophical texts \cite{cohen2014contemporary,sep-theory-bioethics} and collaborating with a domain expert, we have summarized the few most important entities that aid ethical decision-making (see Table \ref{got table}). In this section, we provide ontological unpacking while grounding the used terms in real-world semantics in accordance with ontology development guidelines provided by Guizzardi and Guarino \cite{guizzardi2024explanation}. ApplE explicitly represents entities, or \textit{truthmakers}, that describe an ethical decision instead of treating moral judgement as an atomic classification. Arriving at such an ethical decision thus requires interacting with theoretical and contextual components that lead to interpretable, semantic reasoning. 

ApplE upholds FAIR principles and incorporates interoperability and reusability within its design (see Section \ref{Resource Specifications}). These real-world semantics are thus broadly categorized into two major modules: the Applied Ethics module and the Event Context module.

\begin{table}[!h]\centering
\caption{Glossary of terms for ApplE Ontology.}\label{tab2}
\label{got table}
\begin{tabular}{p{1.5in}p{3in}}
\toprule
\textbf{Term} & \textbf{Definition}\\
\midrule
\multicolumn{2}{c}{\textbf{Applied Ethics Module}}\\ 
\midrule
Ethics & \textit{A discipline concerned with studying what is morally right, wrong, good, or bad.}\\
Applied Ethics & \textit{A discipline concerned with the application of ethics in real-world events.}\\
Applied Ethics Philosophy & \textit{A theory or ideology that stems from normative ethics and is applied to real-world events.}\\
Ethical Principle & \textit{A concept that defends the morality of an action at an intrinsic level.}\\
Ethical Issue & \textit{A situation where an ethical conflict may arise.} \\
\midrule
\multicolumn{2}{c}{\textbf{Event Context Module}}\\ 
\midrule
Context & \textit{The circumstances in which an event occurs.}\\
Event &\textit{Something that happens.} \\
Agent & \textit{An entity capable of doing and receiving actions (eg: Person, Company, Animal, Government, etc.).}\\
Action & \textit{Something that is done.}\\
Consequence & \textit{The outcome of an action.}\\
Moral Intention & \textit{The intention of an agent with regard to a moral action.}\\
\bottomrule
\end{tabular}
\end{table}

\subsection{Applied Ethics Module}

\texttt{AppliedEthics} is a subclass of \texttt{Ethics}, alongside \texttt{MetaEthics} and \texttt{Normative Ethics}. It consists of \texttt{Domain}, such as \texttt{Bioethics} and \texttt{BusinessEthics}. Ethics is a high-level class, which follows a heirarchy: for example, Bioethics is a subclass of Applied Ethics, which is a subclass of Ethics. Since NarrativeEthics is an instance of Bioethics, it is also an instance of class Ethics. Each of these has certain associated \texttt{Applied Ethics Philosophies}, which take into account prior cases to make ethical judgments. 

\begin{figure*}[!ht]
\centering
\includegraphics[width=\textwidth]{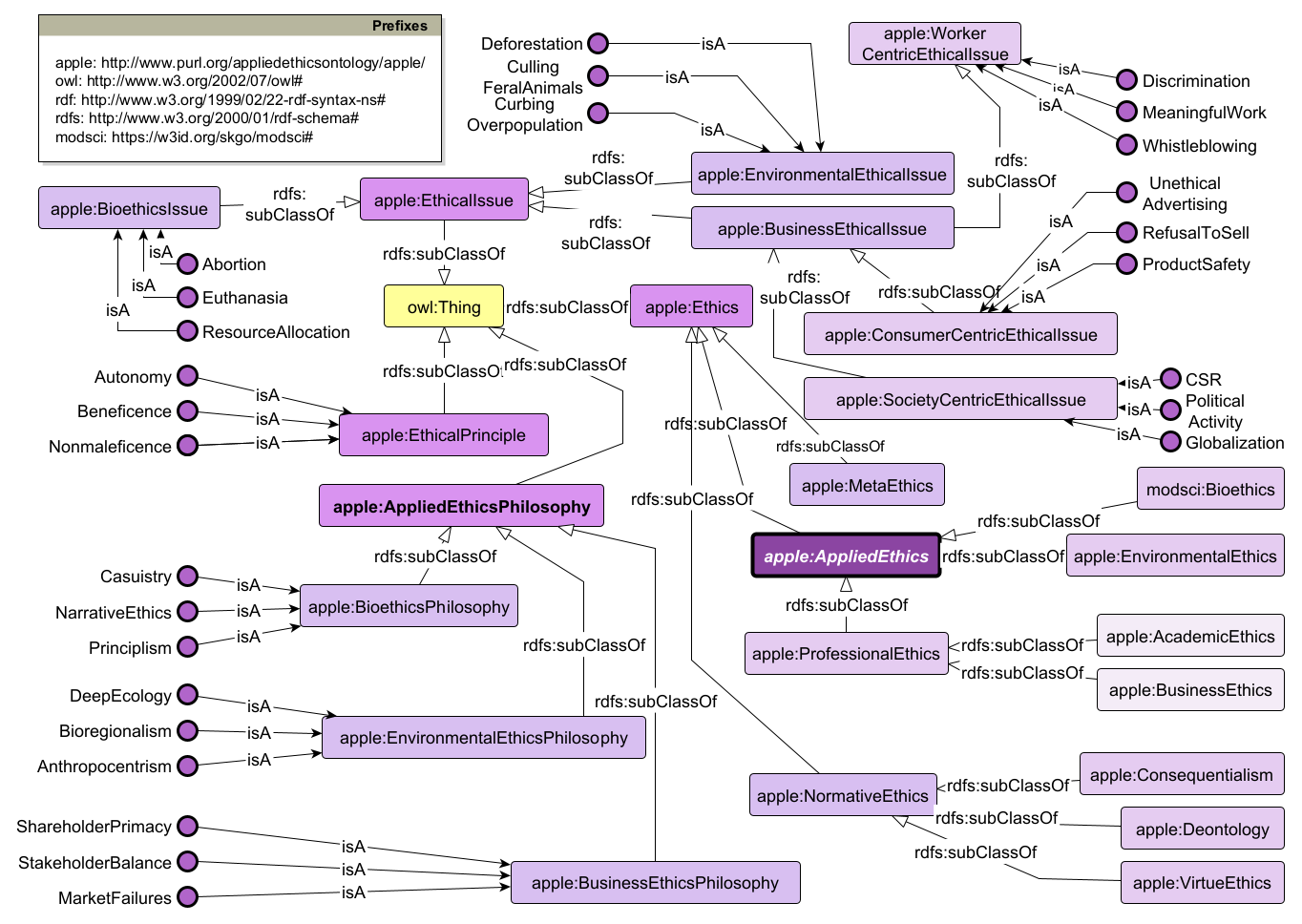}
\caption{A high-level overview that shows the theoretical module of the Applied Ethics ontology. The most important classes have been visualized for the sake of brevity.} \label{fig3}
\end{figure*}

\texttt{NormativeEthics} consists of three subclasses which are important ethical theories. \texttt{Consequentialism} favors \texttt{Consequence} to gauge morality (axiom (5)). \texttt{Deontology} focuses on the intrinsic nature of an \texttt{Action} via a set of rules that may determine if an action is morally right or wrong. \texttt{Action} is linked to one or more \texttt{EthicalPrinciple}, for example, lying is linked to a violation of \texttt{Transparency}, and stealing is a violation of \texttt{nonmaleficence}. 

\vspace{-0.8pt}
Various ethical principles exist, and it would be difficult to provide an accurate taxonomy, especially since individuals value principles differently. However, these principles and moral intention allow us to move away from a purely consequentialist approach, which is not an ideal method to resolve ethical issues \cite{nye2015non}. \texttt{VirtueEthics} describes an ethical action to be performed by a virtuous agent. It is difficult to discretize such virtues in an agent, however, \texttt{MoralIntention} can give insight. Intention provides motivation towards action, and regardless of the consequences or the principles upheld or violated, moral intent is a significant indicator when judging morality.

The applied ethics module also defines a class to capture the \texttt{EthicalIssues} that may arise in a specific domain, which may then be resolved in accordance with the associated \texttt{applied ethics philosophies}. A few axioms from the Applied Ethics module are as follows.
\begin{align}
\centering
    \texttt{AppliedEthics} \sqsubseteq \exists
&\texttt{appliesPhilosophy}. \texttt{Philosophy}  \\
    \texttt{AppliedEthics} \sqsubseteq \exists &\texttt{appliesTo.Domain}\\
    \texttt{EthicalIssue} \sqsubseteq \exists &\texttt{resolvedBy.Philosophy}\\
    \texttt{EthicalPrinciple} \sqsubseteq \exists &\texttt{violatedBy.Action}\\
    \texttt{Consequentialism} \sqsubseteq \exists &\texttt{associatedWith.Consequence}\\
    \texttt{Consequentialism} \sqcap \texttt{D}&\texttt{eontology} \sqcap \texttt{Virtue} \sqsubseteq \bot
\end{align}

Some axioms make explicit the predicates that link two entities together. For example in axioms (1) and (2) applied ethics applies a certain applied ethics philosophy and is applied to a certain domain. Similarly, an ethical principle is resolved by some philosophy (axiom (3)) whereas an ethical principle may be violated by an action (axiom (4)). We may also describe the disjointness of some entities using axioms, as is shown in axiom (6), which indicates that there is nothing common between Consequentialism, Deontology, and Virtue Ethics.  

\subsection{Event Context Module}
\texttt{Context} describes an \texttt{Event} (axiom (7)) which has a \texttt{Domain}, \texttt{TimeEntity}, \texttt{Place}, and the \texttt{Agent} involved. The \texttt{Agent} is assigned a role using \texttt{hasRole} and a moral intention using \texttt{hasMoralIntention}. An agent may be an \texttt{ActiveAgent} or a \texttt{PassiveAgent}. The active agent does or causes an action: \texttt{doesAction(Active Agent, Action)} (axiom (10)); while the passive agent receives the repercussions of the action: \texttt{affects(Action, PassiveAgent)}. An agent's \texttt{Role} may be important to consider in that particular context.

The \texttt{Consequence} is an important consideration and may be characterized by \texttt{SeverityOfConsequence} (axiom (13)), \texttt{UtilityOfConsequence}, and \texttt{Duration-} \texttt{OfConsequence}. It may be difficult to assign these descriptions; however, given the situation at hand, it is easier to make decisions \textit{relatively} between the available courses of action and within a specified context. This has been made evident by the motivating scenario discussed in Subsection \ref{Evaluation}.

An \texttt{action} may be defined as something done while involving certain agents (axiom (8)), leading to some consequences (axiom (9)), and reflecting on some ethical principles. Based on the theoretical factors and contextual information, an action may be indicatively classified as morally right, wrong, or grey. A few axioms from the Event Context module are shown.
\begin{align}
    \texttt{Context} \sqsubseteq \exists &\texttt{describesEvent.Event}\\
    \texttt{Action} \sqsubseteq \exists &\texttt{hasConsequence.Consequence}\\
    \texttt{Action} \sqsubseteq \exists &\texttt{affects.PassiveAgent}\\
    \texttt{ActiveAgent} \sqsubseteq \exists &\texttt{doesAction.Action}\\
    \texttt{Action} \sqsubseteq \exists &\texttt{occursInEvent.Event}\\
    \texttt{MorallyRightAction} \sqsubseteq \exists
    &\texttt{upholdsPrinciple.EthicalPrinciple}\\
    \texttt{Consequence} \sqsubseteq\exists &\texttt{hasSeverity}.\texttt{Severity}\\
    \texttt{Event} \sqsubseteq \exists &\texttt{hasParticipant}.\texttt{Agent}\\
    \texttt{ActiveAgent} \sqsubseteq \exists &\texttt{hasIntention.Intention}
\end{align}

A morally right action, although may also violate some principles, would always uphold at least one ethical principle (axiom (12)). Similarly, a morally wrong action would always violate at least one ethical principle. More explanation on how these classes (morally right, wrong, and grey) are assigned to an instance of the Action class with \textit{SWRL rules} is provided in Section \ref{Evaluation}. The consequence may have some degree of severity, as in axiom (13), the event has some participant who may be an active or passive agent, as in axiom (14), and an active agent may have some moral intention before doing the action, as in axiom (15).

\begin{figure*}[!ht]
\centering
\includegraphics[width=\textwidth]{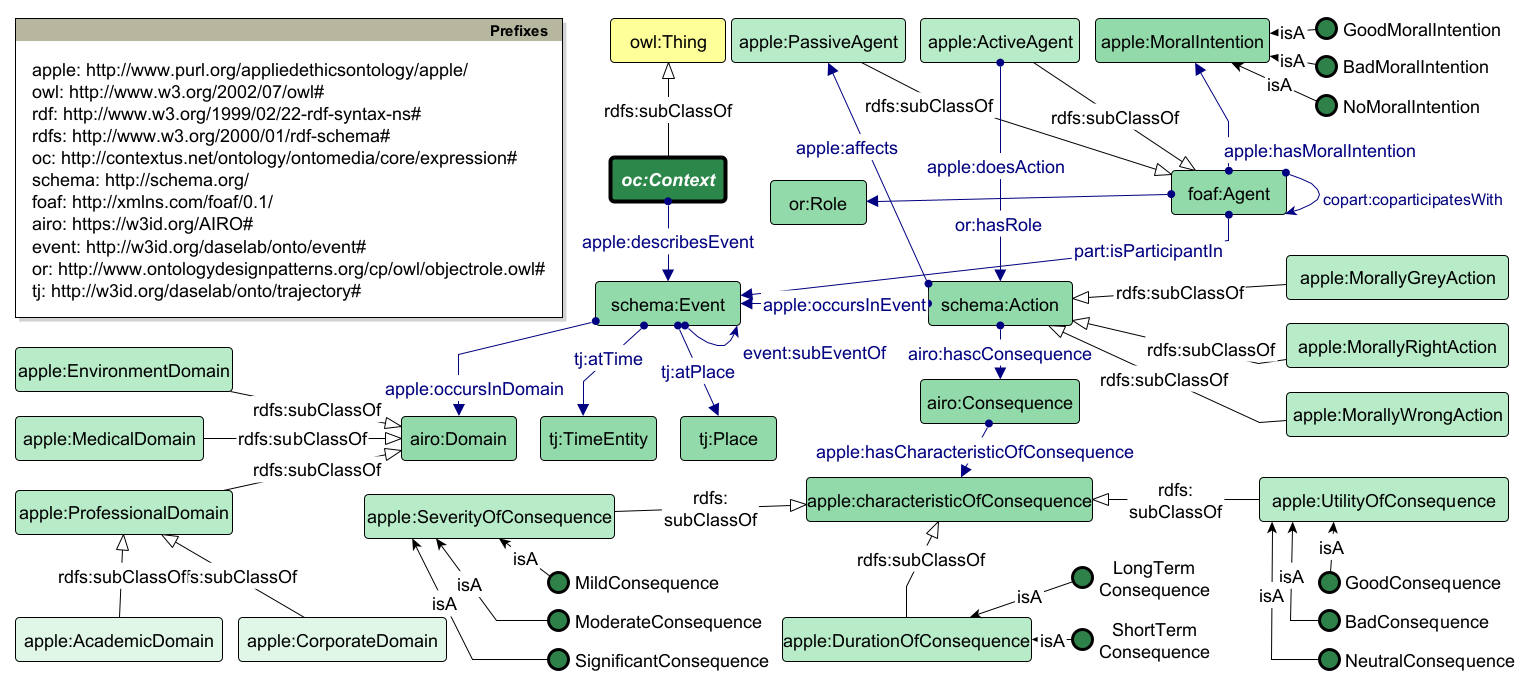}
\caption{An high-level overview that shows the event context module of the Applied Ethics ontology. Some instances and classes have been omitted from this diagram for brevity.} \label{fig4}
\end{figure*}

\section{Resource Specifications}
\label{Resource Specifications} 

\paragraph{FAIR.} As per the FAIR principles \cite{wilkinson2016fair}, the ApplE ontology is findable via a permanent locator and a GitHub repository. These resources will be attached to the final paper upon acceptance for wider dissemination of the ontological resource. It reuses entities from other ontologies and design patterns, which further facilitates interoperability. The naming scheme of the ontology is easily understandable while also being heavily annotated in the ontology documentation. ApplE is open for use as a part or whole under the \href{https://www.apache.org/licenses/LICENSE-2.0}{Apache License 2.0} for interoperability and reusability.
\paragraph{Validation.} Several tools were used validate the ontology apart from a three-fold evaluation. OOPs Pitfall Scanner \cite{poveda2014oops} yielded no pitfalls in the ontology, and Protégé's debugger HermiT (version 1.4.3.456) \cite{glimm2014hermit} deemed ApplE consistent and coherent. A series of twenty competency questions were applied to the model to check the semantic validity of the hierarchy, properties, and axioms.
\paragraph{Documentation.} The technical documentation of this ontology was created using the \href{https://protegewiki.stanford.edu/wiki/OWLDoc}{OWL Documentation} tool in Protégé and will be made available for public access. The documentation will also be made available on GitHub.  
\paragraph{Resource Maintenance.} Any future versions or iterations of the ontology will be updated at the same PURL location in order to remain consistently accessible. Spiral SAMOD's insistence on clear naming, well-documented glossaries, and iterative, modular development, ensures a sustainable development plan. We also take into account newer developments in applied ethics, the inclusion of which would result in yet another development cycle.

\section{Evaluation}
\label{Evaluation}
The SAMOD methodology requires a motivating scenario to evaluate the ontology using three testing methods: model, data, and query testing. This motivating scenario outlines the various classes and relationships used in the ontology via a story that has been adapted to the real world to demonstrate how ApplE models pertinent ethical information in the Bioethics domain, while facilitating reasoning through extensive testing and rules. 

\subsection{In-development Evaluation}
\label{indevelopment eval}
During the development of the ontology, model, data, and query testing is done at every iteration. If the tests fail, it is recommended to go back to a previous milestone and start over. The iterative nature of this approach allowed us to get a clear understanding of how the entities interacted with one another, allowing for a simple but robust ontology. We also utilized SWRL rules \cite{swrl} with the Pellet reasoner \cite{SIRIN200751} to facilitate not only the modeling of the motivating scenario but also support indicative classification of ethics. We first identified a data exemplar or the \textit{motivating scenario} that may be used to demonstrate the three-fold testing process. This scenario describes a morally ambiguous action.

\begin{figure*}[!ht]
\includegraphics[width=\textwidth]{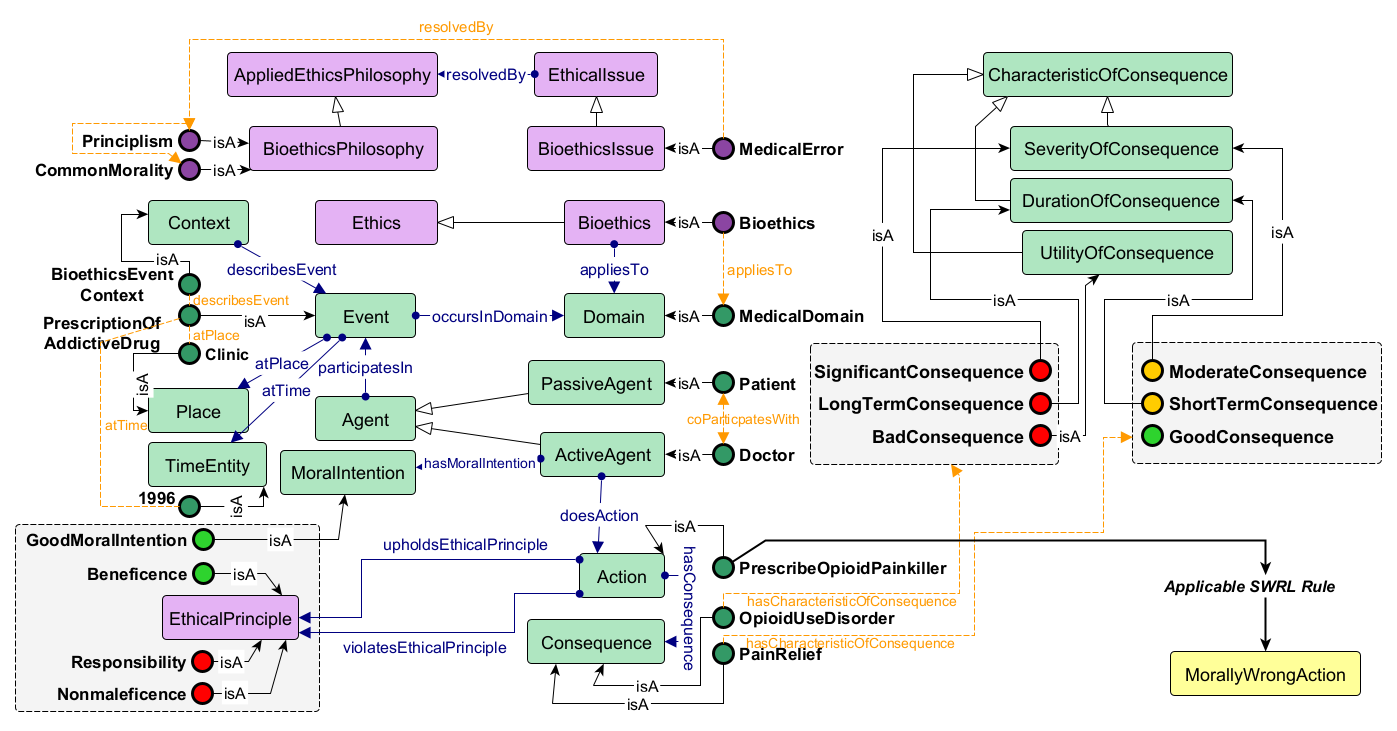}
\caption{A detailed diagram to show how the data exemplar is modeled using the ontology for the Bioethics domain. A similar modeling may be done for any other event of another domain. One of the SWRL rules found a match within the provided data and assigned \texttt{PrescribeOpioidPainkiller} as a Morally Wrong Action accordingly. A few classes and properties have been omitted for the sake of brevity.} \label{use case modeling diagram}
\end{figure*}

\textbf{Motivating Scenario from Bioethics:} With regard to the mass addiction to opioid-based painkillers in the US in the early 2000s \cite{jones2019opioid}: Principlism \cite{beauchamp2004principles} is a bioethics theory that favors justice, nonmaleficence, beneficence, and autonomy. A doctor, with good intentions, prescribed to a teenage patient thirty tablets of OxyContin, a powerful opioid, to relieve post-dental surgery pain. They were not fully aware of its addictive nature when taken over prolonged periods of time, and yet prescribed a large dose of the drug against policy so the patient would not have to come to the clinic repeatedly. However, the continuous intake of the drug led to Opioid Use Disorder (OUD) in the young patient. Given that the good, short-term consequence was mild (relieving pain), but the bad, long-term consequence was significant (addiction to drugs), the doctor was at fault regardless of their good intentions. The doctor violated the principles of responsibility and nonmaleficence while upholding the principle of beneficence. They were responsible for fully knowing the drug before prescribing it and following their duties without breaching any policy. The modeling of this case is visualized in Figure \ref{use case modeling diagram}. The ethical question that needs to be answered is this: 
\emph{Is the doctor's decision to prescribe a large quantity of OxyContin ethically justified despite the intention of relieving the young patient's pain?}

\textbf{Model Testing.} Before we begin to answer this question using ApplE, we must first evaluate the ontology for technical validity. Model testing ensures that the classes, relationships, and instances are semantically intact and serve justified purpose with regard to the domain. The OntoClean methodology \cite{ontocleanguarino2009overview} leads us to a concise ontology with clarity on the rigidity, anti-rigidity, identity, and unity of its entities. We were able to ask questions such as \textit{"What is the difference between an event and an action?"}, \textit{"Why is there a need for a neutral consequence?"}, and \textit{"Is an active agent always active?"}. Furthermore, we checked our ontology via the OOPS Pitfall Scanner \cite{poveda2014oops}, which yielded no pitfalls. We tested our model via queries relying on the HermiT reasoner \cite{hermit} to ensure that there were no unsatisfiable classes and the ontology was behaving the way we expected. Although the object properties have little complexity in terms of hierarchy, we also applied RBox compatibility checks to ensure that they are meaningful to avoid unwanted inferences \cite{keet2018introduction}. 

\textbf{Data Testing.} Now we return to the ethical question about the doctor's action. We first modeled this real-world scenario via ApplE (See Figure \ref{use case modeling diagram}). This demonstrates how ethics theory and event context have been captured to infer morality and evaluate the usefulness of ApplE in terms of real-world compatibility. This answers questions such as, \textit{who acted?} (the Doctor, \texttt{ActiveAgent}), \textit{what did they do?} (prescribed opioid painkiller, \texttt{Action}), and \textit{what were the consequences of their action?} (pain relief and opioid use disorder, \texttt{Consequence}). After incorporating the data instances from the motivating scenario, we created informal competency questions. 

Following the instantiation of the conceptual elements in the ApplE ontology, an applicable SWRL rule independently infers moral classification by aggregating information about the action, such as the characteristics of its consequences, the moral intention of the active agent, and the ethical principles upheld or violated by it. For our bioethics motivating scenario, the applicable rule is as follows,
\begin{quotation}
\textit{ActiveAgent(?ag), hasConsequence(?a, ?c), hasSeverityOfConseque- \newline
nce(?c, ?soc), hasUtilityOfConsequence(?c, ?uoc), hasDurationOfConsequence(?c, ?doc), hasMoralIntention(?ag, ?mi), violatesEthicalPrinciple(?a, ?vep), upholdsEthicalPrinciple(?a, ?uep), hasScore(?soc, ?socs), hasScore(?uoc, ?uocs), hasScore (?doc, ?docs), hasScore(?mi, ?mis), hasScore(?vep,?veps), has Score(?uep, ?ueps), multiply(?prod, ?socs, ?uocs, ?docs), add(?sum, ?prod, ?mis, ?veps, ?ueps), less Than(?sum, 0) -$>$ \textbf{MorallyWrongAction(?a)}}
\end{quotation}
This rule classifies the action \texttt{?a} as a \texttt{MorallyWrongAction}, thus providing an indicative judgement of morality based on semantic theoretical and contextual features of an action as shown in Figure \ref{use case modeling diagram}. ApplE can be used to represent similar scenarios at scale through a domain-specific knowledge graph. The representation of these parameters can be notated mathematically to generate an ethical score using similar rules for resolving ethical dilemmas, as shown in \cite{aijaz2025moral}.

\textbf{Query Testing.} We implemented the competency questions via queries through GraphDB's SPARQL-compliant tool to load the ontology. The competency questions, when formalized using these queries, all provided the expected outcomes. This indicates the correctness and consistency of the ApplE ontology.

\subsection{Post-development Evaluation}
After creating the ApplE ontology, we also evaluated it via some established ontology quality metrics. We chose the semiotic metrics suite as proposed by Burton-Jones et al. \cite{burton2005semiotic}, which requires manual evaluation based on four parameters: \textit{Syntactic, Semantic, Pragmatic, and Social Quality}. The \textit{syntactic} parameters require quality checks with regard to the correctness of syntax. This has been evaluated using the OOPS pitfall scanner and OWL's HermiT. \textit{Semantic} quality evaluates meaningfulness and consistency of the terms that we have used. An advantage of using the Spiral SAMOD methodology was to ensure that the naming of the entities was self-explanatory and clear, hence this is also covered. The \textit{pragmatic} quality ensures that the entities of the ontology are truly domain-specific and justified. Our only deviation from the SAMOD methodology was to seek confirmation from the DE in the second and further iterations of development to satisfy this condition. The \textit{social} quality determines the extent of use. To the best of our knowledge, ApplE is among the first ontologies to explicitly capture Applied Ethics reasoning and concepts. 

\section{Conclusion}
The ApplE ontology models key ethical and contextual information about a morally-ambiguous event to facilitate ethical reasoning. We used the spiral-SAMOD methodology to ensure a standardized ontology and have performed various quality checks. Axioms have been asserted for both the Applied Ethics and Event context modules, taking advantage of property characteristics, subsumptions, and cardinality to explicitly specify relationships between abstract concepts. To demonstrate the applicability of the resource, we modeled a real-world scenario from the bioethics domain. ApplE is FAIR-compliant and is easily accessible for reuse. With this ontology, we aim to provide a mechanism to model ethics of any domain while also facilitating structured ethical reasoning for dilemma resolution. 

\clearpage
\appendix
\section{More Motivating Scenarios}
\begin{figure*}[!ht]
\includegraphics[width=\textwidth]{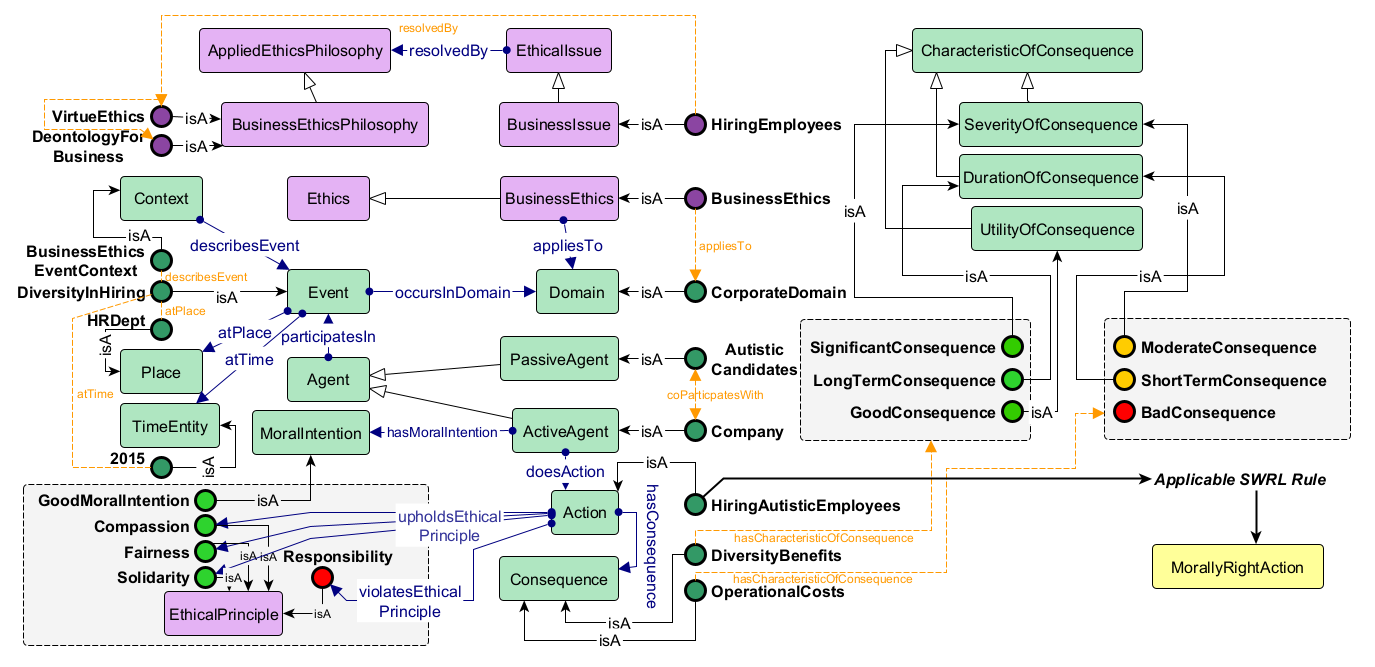}
\caption{A detailed diagram to show how the data exemplar is modeled using the ontology for the Business Ethics domain. The event context and applied ethics module model information about the event, which involves an action that is morally right, in accordance with the provided SWRL rules. A few classes and properties have been omitted for the sake of brevity.} \label{use case modeling diagram 2}
\end{figure*}

\paragraph{\textbf{Motivating Scenario from Business Ethics:}} Microsoft’s Autism Hiring Program \cite{microsoftautism} focuses on hiring autistic individuals, utilizing their talents, and improving their standard interview processes for inclusivity. They aim to provide these hires with accommodations specific to their needs in the workplace \cite{morris2015understanding}. Virtue ethics and deontology are often cited for use in making moral decisions in business or professional scenarios. Being inclusive of all members of society when providing meaningful work, focusing on skill over operational costs, and actively accommodating the lesser represented are all aspects portrayed by a virtuous moral entity. Furthermore, such an entity, in this case, Microsoft, upholds various ethical principles such as compassion, fairness, solidarity, justice, and integrity \cite{grimani2020fostering}. However, one may argue that hiring candidates with more needs at the workplace would not be in the best interests of the stakeholders, as it increases hiring costs and also requires more support to these employees over others, causing an imbalance of consideration at the workplace. In such cases, Microsoft would be compromising its responsibility to its stakeholders; the shareholders and its other employees. A counterargument to this would lie in the consequences of these actions; not only would hiring autistic candidates bring benefits of inclusivity, but it would also tap into previously unrecognized talent and build the company's reputation. These aspects make it preferable, over time, for the stakeholders in the company as well. Overall, the act of hiring candidates who are on the autistic spectrum would be a morally right one, as this empowers underrepresented groups, reduces bias, and ensures equitable job offers \cite{woods2021hiring}. This case has been visualized in figure \ref{use case modeling diagram 2}.

\begin{figure*}[ht]
\includegraphics[width=\textwidth]{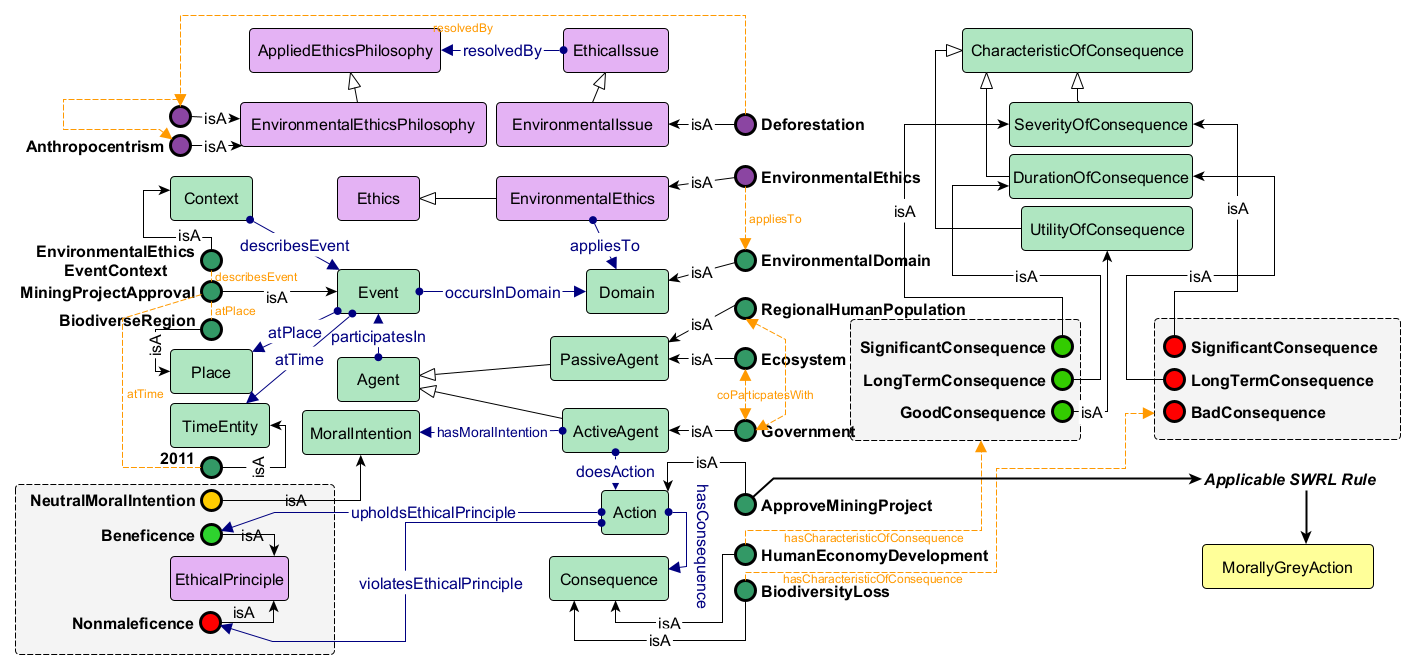}
\caption{A detailed diagram to show how the data exemplar is modeled using the ontology for the Environmental Ethics domain. The event context and applied ethics module model information about the event, which involves an action that is morally grey, in accordance with the provided SWRL rules. A few classes and properties have been omitted for the sake of brevity.} \label{use case modeling diagram 3}
\end{figure*}

\paragraph{\textbf{Motivating Scenario from Environmental Ethics:}} This case pertains to the deforestation and developmental pressures on India’s Western Ghats, a UNESCO World Heritage site renowned for its rich biodiversity and high level of endemism \cite{UNESCO2012}. For over the past decade, the Western Ghats, a tropical forest area with significant biodiversity, face significant threats from deforestation for agricultural expansion, mining, and infrastructure projects. However, there is an increasing human population in this region which must be accommodated, and their livelihoods must also be considered. Furthermore, the Indian government also prioritizes eco-tourism, promotion of bio-mass and agriculture-based industries, and regulated transportation and infrastructure \cite{nagarajan2015appraising}. All these would bring about much improvement in the lives of the people living in the Western Ghats. In environmental ethics, anthropocentrism is an approach to resolving such environmental concerns, by prioritizing human interests and considering all other beings to be means to human ends \cite{probyn2018anthropocentrism}. As per domain experts, this theory consideration creates a morally grey dilemma, where on one hand, economic development in these areas provides job opportunities, improves infrastructure, and raises the standard of living for local communities. On the other, such actions lead to habitat destruction, biodiversity loss, and potentially irreversible environmental damage, impacting the ecosystem and future generations. By adhering to an anthropocentric approach, the government upholds the principle of beneficence toward human populations in the long term, as development may offer prolonged socio-economic benefits. However, this perspective neglects the principle of nonmaleficence regarding the environment, as it allows for ecological harm to fulfill human needs. Consequently, while the anthropocentric approach rationalizes development to aid human welfare, it fails to consider the intrinsic value of the Western Ghats’ unique ecosystems, ultimately making this case morally complex\footnote{This example in particular poses a certain caveat. The SWRL rule that indicates morally grey actions requires an equivalence of both negative and positive repercussions, and principles upheld and violated, while also requiring a neutral moral intention. This representation of a morally grey situation may be considered too rigid, and may exclude cases where one scenario is only slightly better than the other. This is why, although the ApplE ontology does an excellent job of capturing the pertinent nuances of a complex scenario such as this by adding more details via data instances, the SWRL rules are expected to work better where the ethical decisions are more inclined to one side than the other. In order to resolve more difficult ethical dilemmas, a learning model or a case-based reasoner may be used in conjunction with the ApplE ontology.}. This case has been visualized in figure \ref{use case modeling diagram 3}.

\section{Discussion}
\label{discussion}
In this section, we discuss some perspectives that relate to our proposition and may provide additional context for readers. We provide a comparison of using inference techniques for capturing abstract concepts as opposed to learning them. We also highlight a philosophical overview of the development of such AI systems, which may leverage the use of resources such as ApplE. We anticipate and discuss the limitations of our resource, along with how these findings may fuel future research in this space.

\subsection{Inferring versus Learning Ethical Concepts}
\label{learn}
Learning concepts from large amounts of data using algorithms has various advantages. Oftentimes, connectionist models that are deep and extensive enough are able to find insights that have been overlooked by human investigators. For various prediction, classification, and diagnosis applications, machine learning models are sought after for state-of-the-art results. However, in the context of a subject like ethics, machine learning alone will not suffice to provide results.

This understanding was reached after an extensive review of contributions that mainly used learning algorithms to provide ethical dilemma resolution in narrow domains \cite{zhong2025computational}. Although these models were able to provide insights into the morality of various AI decisions and actions, their results sometimes vary, even in similar cases. We notice this in the behavior of large language models, which provide hallucinatory, inconsistent, and unreliable information \cite{huang2025survey}. Since these models are usually black boxes, it is difficult to judge where the decision-making deviated without looking at small meaningful perturbations \cite{fong2017interpretable}.

Even if we can point out where the deviation occurred, the \textit{why} oftentimes remains a mystery, requiring further magnification. This is implausible for the consideration of moral sanctity in a situation. To consider a rhetorical example: in two very similar cases of the trolley problem \cite{shallow2011trolley}, either two passengers could be saved or three pedestrians. The characteristics of the passengers remained the same in both cases, but a slight change in one of the pedestrians causes all three to be sacrificed by an autonomous steering system. \textit{Why did the model make this decision?} Ideally, based on analogical reasoning, two vastly similar cases should lead to very similar decisions. Explainability is imperative in ethics modeling, and it is difficult to find purely connectionist models that are able to provide adequate\footnote{By adequate, we imply that an explanation for the resolution of an ethical dilemma would need to provide information on all the factors involved and pinpoint exactly where the deviation occurred and why. Current explanation models do not provide such details, especially when applied to the ethics domain. They provide a general idea of the degrees of contribution to the results for each of the important features.} explanations. 

Other issues with using machine learning algorithms include their reliance on large amounts of data. This leads to the reflection of inherent biases, inaccuracies, and inconsistencies in the data. When working on modeling ethics, this issue is perhaps the most concerning. The very inclusion of bias in a system that aims to provide ethical resolution of dilemmas is unethical at the premise. Even large, extensive representative data sets may not be able to address the problem of bias. Bias may also result from decisions that follow just procedures but may lead to unfair outcomes for certain groups. This may be due to a lack of inclusion, unfair representation, or negative prejudice against a certain community in the dataset \cite{barocas2016big}. Extensive techniques are required to reduce bias in the data, and they still may not eliminate it entirely. Furthermore, data for such models are often self-reported and thus may not reflect reality; rather, they report opinions and perspectives, which may not always be useful. Self-reporting involves a projection of what an agent conceives to be an ideal action in the context. This may still be an outcome of deep-seated prejudices that guide their belief. Ethical reasoning involves engagement with not only the primary motivation for an action, but also justification, which involves second-order ethical reasoning.
Using knowledge representation techniques, such as an ontology which may be reasoned over, provides some rigidity in ethical decision-making. This would allow a user to trace back to why a decision was made and where it may have deviated from other similar cases. However, such a rigid system may still not be feasible to work on highly grey areas of morality. Thus, an inference model in conjunction with a learning model has the best chance of building a moral machine. 
Moral realism \cite{railton2020ethical} conceives a moral agent akin to a child who, based on their experiential world and demands of morality, tends to revise their beliefs before solidifying them with age. In a similar vein, neurosymbolic models can triangulate meaningful, consistent results to become a \textit{moral agent}. A knowledge system acts as a constraint on autonomous moral reasoning; the violation of which constitutes a violation of the inherent duty of the machine to uphold ethical principles. A learning system can also track the evolving nature of ethical principles and norms and respond to novel challenges.
The ApplE ontology is a resource contribution that is rooted in this philosophy, and builds on these questions lying at the intersection of ethics and AI. Various authors such as Savulescu and Malsen \cite{savulescu2015moral}, Bro{\.z}ek and Janik \cite{brozek2019can}, Pana \cite{pana2006artificial}, and Liao \cite{liao2020ethics,railton2020ethical} have penned their thoughts on artificial morality and the capacity for ethical decision-making in AI systems. With reference to these and more, we have formulated an Applied Ethics ontology that is built based on the way human beings make decisions in the presence of ethical ambiguity. However, as humans are predisposed to their background and upbringing, which, for the most part, dictates how they make their decisions. AI systems, which may be considered to be in a Lockean blank state \cite{uzgalis2001john}, do not have to carry those biases and preconditioning to make moral decisions. 
Yet this lack of grounding also calls for caution, as it equally enables the possibility of repeating humanity's errors. Just as our capacity of natality makes it possible to see the world with fresh eyes \cite{arendt2022human}, AI too may learn from the \textit{data} provided to it. However, it also needs \textit{knowledge} to make sense of the data it holds. Thus, ApplE does not aim to make decisions on behalf of autonomous agents, rather it aims to enable them to make ethically-informed ones.
This contribution also fuels both technological and metaethics research, as we can run and test neurosymbolic models built upon this resource to truly gauge (in real-time) the applicability of specific theories (for example, principlism from our bioethics example) to some domain-specific use cases. Not only would this improve the technology deployed, but also our own understanding of what works best in certain situations, aided by the study of casuistry \cite{sep-theory-bioethics}.

\subsection{Limitations and Future Work}
\label{limit}
Applied ethics requires capturing context as well as ethics theories. The authors of this ontology acknowledge the fact that the entities considered are not exhaustive of all that ethicists may consider when making a decision on a matter of moral ambiguity. However, in order to discretize the ethical decision-making process, we had to select only those factors that contribute most to the final decision. There may, however, be scenarios where an aspect that could change the decision was not considered, as it was not included in the ontological representation of the scenario. We do not readily anticipate this, but it is possible. 
There are also some researchers who counter the development of \textit{general} resources. Raji et al., mention the issues with developing such general benchmarks, such as how their performances are contextualized from any domain and may lead to inappropriate use \cite{raji2021ai}. Gebru and Denton state that in order to cater to a varied user base, developers \textit{must} provide many different solutions, rather than one scalable solution \cite{gebru2024beyond}. The development of the ApplE ontology should not be considered entirely general, however, and must instead be considered a top-down approach. We have developed from a taxonomy of applied ethics and event context an ontology that also considers the applicability of various domains. The ApplE ontology may be used to model specific cases from the applied ethics domains, and may also be reused in its parts or as a whole to build a domain-specific ethics ontology.
In addition to this, there is scope to use the ontology along with some learning algorithms and more sophisticated reasoning methods to apply ethical dilemma resolution and case-based reasoning, as discussed in Subsection \ref{learn}. ApplE can adequately \textit{represent} ethics during an event, and using SWRL rules present an \textit{indication} towards whether or not the action is morally right, wrong, or grey.
For fuzzy interpretations of grey areas, the ApplE ontology may be combined with a learning model to provide better results. Such algorithmic models may also facilitate the automatic selection of theories and grey area thresholds. ApplE may also be used to create modular ontologies for specific applied ethics domains. It may also be used as a benchmark for knowledge graphs which may be queried to provide a strong dilemma-resolving, inference-based models.   
Ethical reasoning in human agents constitutes the possibility of counterfactual reasoning linked to complex emotions \cite{migliore2014counterfactual}. Human agents are not only able to generate ethical responses based on an assessment of the facts, but they are also able to assess alternate "what-ifs" when a situation changes. In other words, human agents have the capacity of reflection and also meta-reflection, not only with regard to the rightness of an action, but also on the dynamic peculiarities of context. It is a challenge that artificial agents have faced and not always adequately responded to because all the particularities cannot be learned in advance. Much of it requires participation and paying attention. In machines, the absence of accountability and explainability is bound to have catastrophic consequences. We anticipate that the initial use of ApplE will be in domains where norms have been adequately established, and probability of high-risk counterfactuals is less.

\section{ApplE's SAMOD Methodology Checklist}
\begin{enumerate}
    \item Domain Experts (DE) \checkmark
    \item Ontology Engineers (OEs) \checkmark
    \item Quick iterative cycles to produce \textit{modelets}. \checkmark
    \item Self-explanatory naming scheme for entities and relationships \checkmark 
    \item Glossary of terms \checkmark 
    \item T-Box/A-Box consistency \checkmark
    \item Data exemplars/Motivating Scenarios \checkmark 
    \item Three-fold testing (model, data, and query testing) \checkmark 
    \item Modular development \checkmark
    \item Reusing ODPs \checkmark
    \item Top-down development \checkmark
    \item Documentation \checkmark
\end{enumerate}

\section{Competency Questions}
\label{CompetencyQuestions}
These tables list some selected CQs used to model and data test ApplE. The basic queries were accomplished via Description Logic (see table \ref{tab4} and \ref{tab5}), and the more complex ones used SPARQL (see tables \ref{tab6} and \ref{tab7}).

\begin{table*}[p]
\centering
\caption{Competency Questions with Description Logic - Model Queries}
\label{tab4}
\makebox[\textwidth][c]{
\begin{tabular}{p{0.5in}p{1.7in}p{2in}p{1.7in}}  
\toprule
Identifier & CQ & DL Query & Results \\
\midrule
CQ1 & What are the main branches of Ethics and its subclasses? & DL:Ethics & Results: AppliedEthics, Bioethics, BusinessEthics, Consequentialism, Deontology, EnvironmentalEthics, EthicsOfCare, LegalEthics, MediaEthics, MetaEthics, NormativeEthics, PoliticalEthics, ProfessionalEthics, TechnologicalEthics, VirtueEthics \\
CQ2 & What are the characteristics of a consequence of an Action? & CharacteristicOfConsequence & DurationOfConsequence, SeverityOfConsequence, UtilityOfConsequence \\
CQ3 & Are there any common issues that exist in both Bioethics and BusinessEthics? & BioethicsIssue and BusinessEthicalIssue & Consent \\
CQ4 & Which domain adheres to the applying Bioethics? & Domain and adheresTo some Bioethics & MedicalDomain \\
CQ5 & Which kinds of applied ethics adhere to the professional domain? & AppliedEthics and appliedTo some ProfessionalDomain & AcademicEthicsBusinessEthics \\
CQ6 & What are the different Philosophies applied to Environmental Ethics? & EnvironmentalEthicsPhilosophy & Animism, Anthropocentrism, Bioregionalism, DeepEcology, Disenchantment, Feminism, SocialEcology \\
CQ7 & Which applied ethics apply the philosophy of Feminism? & AppliedEthicsPhilosophy and \{Feminism\} & Bioethics, EnvironmentalEthics \\
CQ8 & Which Agent does an action? & foaf:Agent and doesAction some Action & ActiveAgent\\
CQ9 & Which branch of Normative ethics is associated with Moral Intention? & NormativeEthics and associatedWith some MoralIntention & Virtue Ethics \\
CQ10 & What kinds of issues are resolved by Bioethics Philosophies? & resolvedBy some BioethicsPhilosophy & Abortion, ClinicalResearch, Consent, DoNotResuscitate, Euthanasia, Misdiagnosis, ResourceAllocation \\
CQ11 & Who is affected by an Action? & inverse affects some Action & PassiveAgent \\
CQ12 & What kind of action may both uphold and violate some ethical principles? & Action and (upholdsEthicalPrinciples some) and (violatesEthicalPrinciples some) & MorallyGreyAction \\
CQ13 & Which Business Ethical Issues are resolved by the stakeholder balance philosophy? & BusinessEthicalIssue and (resolvedBy some \{StakeholderBalance\}) & Consent, DiscriminationInEmployment, FiringWithoutJustCause, PriceDiscrimination, ProductSafety, UnreasonableCompensation \\
CQ14 & Which Applied ethics philosophy may help to resolve the issue of Deforestation? & AppliedEthicsPhilosophy and (resolves some \{Deforestation\}) & DeepEcology \\
\bottomrule
\end{tabular}}
\end{table*}

\begin{table*}
\centering
\caption{Competency Questions with Description Logic - Data Queries}
\label{tab5}
\makebox[\textwidth][c]{
\begin{tabular}{p{0.5in}p{1.7in}p{2in}p{1.7in}}  
\toprule
Identifier & CQ & DL Query & Results \\
\midrule
CQ15 & Who are the participants in the event of prescribing an addictive drug? & isParticipantIn some \{PrescriptionOfAddictiveDrug\} & Doctor, Patient \\
CQ16 & Which active agent prescribes the opioid painkiller? & ActiveAgent and doesAction some \{PrescribeOpioidPainkiller & Doctor \\
CQ17 & What are the consequences of the action done by the Doctor? & Consequence and inverse hasConsequence some (Action and inverse doesAction some \{Doctor\}) & OpioidUseDisorder, PainRelief \\
CQ18 & What are the characteristics of the consequence: Opioid use disorder?
& CharacteristicOfConsequence and inverse hasCharacteristicOfConsequence some \{OpioidUseDisorder\}
& BadConsequence, LongTermConsequence, SignificantConsequence \\
CQ19 & What ethical principles are violated by the action of the doctor?
& inverse violatesEthicalPrinciple some (Action and inverse doesAction some \{Doctor\}) & Nonmaleficence, Responsibility \\
CQ20 & What is the severity of the consequence pain relief? & SeverityOfConsequence and inverse hasSeverityOfConsequence some \{PainRelief\} & MildConsequence \\
\bottomrule
\end{tabular}
}
\end{table*}

\begin{table*}
\centering
\caption{Competency Questions with SPARQL - Model Queries}
\label{tab6}
\makebox[\textwidth][c]{
\begin{tabular}{p{0.5in}p{1.3in}p{2.4in}p{1.7in}}  
\toprule
Identifier & CQ & SPARQL Query & Results \\
\midrule
CQ21 & Feminism is a type of which applied ethics philosophy? & 
SELECT DISTINCT ?principle\}\newline
?principle rdfs:subClassOf \newline apple:AppliedEthicsPhilosophy.\newline
    ?f a ?principle.\newline
    FILTER CONTAINS(str(?f), "Feminism").\}\newline
& BioethicsPhilosophy, EnvironmentalEthicsPhilosophy\\
CQ22 & Which Business Ethical Issues are resolved by the stakeholder balance philosophy? & SELECT ?ethicalIssue\newline
WHERE \{\newline
  ?ethicalIssue apple:resolvedBy apple:StakeholderBalance.
\} & Consent, DiscriminationInEmployment, FiringWithoutJustCause, PriceDiscrimination, ProductSafety, UnreasonableCompensation\\
CQ23 & Which environmental ethics philosophy may resolve the issue of deforestation? & SELECT ?philosophy\newline
WHERE \{\newline
  ?philosophy a \newline apple:EnvironmentalEthicsPhilosophy .\newline
  ?philosophy \newline apple:resolves apple:Deforestation .\}\newline
& DeepEcology \\
\bottomrule
\end{tabular}
}
\end{table*}

\begin{table*}
\centering
\caption{Competency Questions with SPARQL - Data Queries}
\label{tab7}
\makebox[\textwidth][c]{
\begin{tabular}{p{0.5in}p{1.3in}p{3.3in}p{1in}}  
\toprule    
Identifier & CQ & SPARQL Query & Results \\
\midrule
CQ24 & List all actions where an active agent and passive agents are coparticipants. & SELECT DISTINCT ?action ?activeAgent ?passiveAgent\newline
WHERE \{ ?action apple:affects ?passiveAgent . ?action :doneBy ?activeAgent. ?activeAgent coparticipation:coparticipatesWith ?passiveAgent.\} & PrescriptionOf- AddictiveDrug \\
CQ25 & What are the characteristics of the consequence "Pain Relief"? & SELECT ?consequence ?severity ?utility ?duration\newline
WHERE \{ ?consequence a AIRO:Consequence . ?consequence :hasSeverityOfConsequence ?severity. ?consequence :hasUtilityOfConsequence ?utility. ?consequence :hasDurationOfConsequence ?duration. \newline
FILTER contains(str(?consequence), "PainRelief") \} & PainRelief, MildConsequence, GoodConsequence, ShortTermConsequence \\
CQ26 & List the ethical principles upheld by agents who have performed actions in at least one event, and count how many events they participated in. & SELECT ?agent ?upheldPrinciple (COUNT(DISTINCT ?event) AS ?eventCount)\newline
WHERE \{ ?action :doneBy ?agent . ?action apple:occursInEvent ?event . ?action apple:upholdsEthicalPrinciple ?upheldPrinciple .\}\newline
GROUP BY ?agent ?upheldPrinciple\newline
HAVING (COUNT(DISTINCT ?event) > 1)\newline
ORDER BY DESC(?eventCount) & Doctor, Beneficence, 1 \\
CQ27 & What are the agents who perform actions in an event and those who do not perform any actions in that event? What is the time and place of that event? & 
SELECT ?agent ?event ?action ?time ?place \newline
WHERE \{  ?agent a foaf:Agent . ?event trajectory:atTime ?time.  ?event trajectory:atPlace ?place. \newline 
OPTIONAL \{  ?action :doneBy ?agent . ?action apple:occursInEvent ?event . \} \} \newline
GROUP BY ?agent ?event ?action ?time ?place \newline
ORDER BY DESC(?time) \newline & Doctor;Patient, PrescriptionOfAddictiveDrug, 1996, Clinic \\
CQ28 & Which active agents have participated in at least one action, and what ethical principles do these actions uphold or violate? & 
SELECT ?agent (COUNT(?action) AS ?actionCount) ?action ?upheldPrinciple ?violatedPrinciple \newline
WHERE \{ ?action :doneBy ?agent . \newline
OPTIONAL \{ ?action apple:upholdsEthicalPrinciple ?upheldPrinciple . \} \newline
OPTIONAL \{ ?action apple:violatesEthicalPrinciple ?violatedPrinciple . \}\}\newline
GROUP BY ?agent ?action ?upheldPrinciple ?violatedPrincipleHAVING (COUNT(?action) > 0) & Doctor, 1, PrescriptionOfAddictiveDrug, Beneficence, Responsibility;Nonmaleficence \\
CQ29 & Which actions were done by a Doctor and do not violate any ethical principles? & SELECT ?action ?principle \newline
WHERE \{ ?action a schema:Action . \newline
?action applehash:doneBy applehash:Doctor .  \newline
FILTER NOT EXISTS \{ ?action apple:violatesEthicalPrinciple ?principle .\}\} & None \\
CQ30 & For each ethical principle, how many distinct actions uphold it, and how many violate it? & 
SELECT ?principle ?upholdingAction ?violatingAction (COUNT(DISTINCT ?upholdingAction) \newline
AS ?upheldCount) (COUNT(DISTINCT ?violatingAction) AS ?violatedCount) \newline
WHERE \{ \{ ?upholdingAction apple:upholdsEthicalPrinciple ?principle . \} \newline
UNION \{ ?violatingAction apple:violatesEthicalPrinciples ?principle . \}\} \newline
GROUP BY ?principle ?upholdingAction ?violatingAction & Beneficence, PrescribeOpioidPainkiller, none, 1, 0 \\ \bottomrule
\end{tabular}
}
\end{table*}

\section{Namesapces of reused ODPs and ontologies in ApplE}

See Table \ref{tab1}.
\begin{table}[!ht]
\centering
\caption{Namespaces of reused ODPs and ontologies in ApplE.}
\label{tab1}
\begin{tabular}{p{1.2in}p{3.5in}}
\toprule
\textbf{Class/Property} & \textbf{Namespace} \\
\midrule
Consequence, & airo: https://w3id.org/AIRO \\ 
hasConsequence, & \\
Action, Domain & \\

coParticipatesWith & copart:http://www.ontologydesignpatterns.org/ \\
 & cp/owl/coparticipation.owl \\

subEventOf & event: http://w3id.org/daselab/onto/
event \\

Context & ex: http://contextus.net/ontology/ontomedia/ \\
 & core/expression \\

Agent & foaf: http://xmlns.com/foaf/0.1/ \\

Bioethics & modsci: https://w3id.org/skgo/modsci \\

Role, hasRole & or: http://www.ontologydesignpatterns.org/ \\
 & cp/owl/objectrole.owl \\
 
hasParticipant, & part:http://www.ontologydesignpatterns.org/ \\
isParticipantIn & cp/owl/participation.owl \\
 
Event & schema: http://schema.org/\\

hasBeginning/End & time: http://www.w3.org/2006/time\\

doesAction & traffic:http://www.sensormeasurement.appspot.com/\\
 & ont/transport/traffic \\

Time, atTime, & tj:http://w3id.org/daselab/onto/trajectory \\
Place, atPlace  & \\
\bottomrule
\end{tabular}
\end{table}

\bibliographystyle{splncs04}
\bibliography{refer}

\end{document}